\documentclass{PoS}

\title{Dark Energy density in Split SUSY models inspired by degenerate vacua}

\ShortTitle{Dark Energy density in Split SUSY models inspired by degenerate vacua}

\author{Colin Froggatt\\
        Glasgow University\\
        E-mail: \email{c.froggatt@physics.gla.ac.uk}}

\author{\speaker{Roman Nevzorov}%
         \thanks{On leave of absence from the Theory Department, ITEP, Moscow, Russia.}\\
        University of Hawaii\\
        E-mail: \email{nevzorov@phys.hawaii.edu}}

\author{Holger Bech Nielsen\\
        The Niels Bohr Institute\\        
        E-mail: \email{hbech@nbi.dk}}

\abstract{
It is well known that global symmetries protect local supersymmetry and a zero value 
for the cosmological constant in no--scale supergravity. The breakdown of these 
symmetries, which ensure the vanishing of the vacuum energy density, results in a 
set of degenerate vacua with broken and unbroken supersymmetry leading to the natural 
realisation of the multiple point principle (MPP). Assuming the degeneracy of
vacua with broken and unbroken SUSY in the hidden sector we estimate the value of 
the cosmological constant. We argue that the observed value of the dark energy density 
can be reproduced in the split-SUSY scenario if the SUSY breaking scale is of the order 
of $10^{10}\,\mbox{GeV}$.}

\FullConference{35th International Conference of High Energy Physics\\
                 July 22-28, 2010\\
                 Paris, France}

\begin{document}

\section{No-scale supergravity and the multiple point principle}

Recent observations indicate that 70\%-73\% of the energy density of the Universe 
exists in the form of dark energy. This tiny vacuum energy density (the cosmological 
constant) $\Lambda \sim 10^{-123}M_{Pl}^4 \sim 10^{-55} M_Z^4$ is responsible for 
the accelerated expansion of the Universe. In the standard model (SM) the cosmological 
constant is expected to be many orders of magnitude larger than the observed vacuum 
energy density. An exact global supersymmetry (SUSY) ensures zero value for the 
vacuum energy density. However the breakdown of SUSY induces a huge and positive 
contribution to the cosmological constant of order $M_{S}^4$, where the SUSY breaking 
scale $M_{S}\gg 100\,\mbox{GeV}$.

In general the vacuum energy density in $(N=1)$ supergravity (SUGRA) models 
is huge and negative $\Lambda\sim -m_{3/2}^2 M_{Pl}^2$, where $m_{3/2}$ is 
the gravitino mass. The situation changes dramatically in no-scale supergravity 
where the invariance of the Lagrangian under imaginary translations and 
dilatations results in the vanishing of the vacuum energy density. Unfortunately 
these global symmetries also protect supersymmetry which has to be broken in 
any phenomenologically acceptable theory. The breakdown of dilatation 
invariance does not necessarily result in a non--zero vacuum energy 
density \cite{1}--\cite{2}. This happens if the dilatation invariance is 
broken in the superpotential of the hidden sector only. The hidden sector of 
the simplest SUGRA model of this type involves two singlet superfields, 
$T$ and $z$. The invariance under the global symmetry transformations 
constrains the K$\ddot{a}$hler potential and superpotential of the hidden 
sector, which can be written in the following form \cite{1}:
\begin{equation}
\hat{K}=-3\ln\Biggl[T+\overline{T}-|z|^2 \Biggr]\,,\qquad 
\hat{W}(z)=\kappa\Biggl(z^3+\mu_0 z^2+\sum_{n=4}^{\infty}c_n z^n \Biggr)\,.
\label{1}
\end{equation}
Here we use standard supergravity mass units: $\frac{M_{Pl}}{\sqrt{8\pi}}=1$.
The bilinear mass term for the superfield $z$ and the higher order terms $c_n z^n$ 
in the superpotential $\hat{W}(z)$ spoil the dilatation invariance. However the
SUGRA scalar potential of the hidden sector remains positive definite in the 
considered model
\begin{equation}
V(T,\, z)=\frac{1}{3(T+\overline{T}-|z|^2)^2}\biggl|\frac{\partial \hat{W}(z)}{\partial z}\biggr|^2\,,
\label{2}
\end{equation}
so that the vacuum energy density vanishes near its global minima. In the simplest case 
when $c_n=0$, the scalar potential (\ref{2}) has two extremum points at $z=0$ and 
$z=-\frac{2\mu_0}{3}$. In the first vacuum, where $z=-\frac{2\mu_0}{3}$, local supersymmetry 
is broken and the gravitino gains a non--zero mass. In the second minimum, the vacuum 
expectation value of the superfield $z$ and the gravitino mass vanish.

Thus the considered breakdown of dilatation invariance leads to a natural realisation of 
the multiple point principle (MPP). The MPP postulates the existence of many phases allowed
by a given theory having the same energy density \cite{3}--\cite{31}. In SUGRA 
models of the above type there is a vacuum in which the low--energy limit of the considered 
theory is described by a pure supersymmetric model in flat Minkowski space. According to the 
MPP this vacuum and the physical one in which we live must be degenerate. Such a second 
vacuum is only realised if the SUGRA scalar potential has a minimum where $m_{3/2}=0$ which 
normally requires an extra fine-tuning \cite{4}. In the SUGRA model considered above the 
MPP conditions are fulfilled automatically without any extra fine-tuning at the tree--level.
We now assume the existence of a phenomenologically viable model of this type having the 
physical vacuum degenerate with a second supersymmetric flat vacuum.

\section{The value of the dark energy density}
Since the vacuum energy density of supersymmetric states in flat Minkowski space 
is just zero and all vacua in the MPP inspired SUGRA models are degenerate, the
cosmological constant problem in the physical vacuum
is thereby solved to first approximation by our 
assumption. However non--perturbative effects in the observable sector 
can give rise to the breakdown of SUSY in the supersymmetric state
(phase) at low energies. Then the MPP assumption implies that the physical 
phase in which local supersymmetry is broken in the hidden sector has  
the same energy density as the phase where supersymmetry breakdown takes 
place non-perturbatively in the observable sector.

In order to simplify our analysis we restrict our consideration by taking the 
simplest canonical form for the gauge kinetic functions $f_a(T,\, z)\simeq const$.
Due to the mild dependence of $f_a(T,\, z)$ on $T$ and $z$ the gauginos of mass 
$M_a$ are typically substantially lighter than the scalar 
particles of mass $m_{\alpha}$ in the considered 
SUSY models, i.e. $M_a\ll m_{\alpha}\sim m_{3/2}$. Such a hierarchical structure 
of the particle spectrum naturally appears in Split Supersymmetry models.

If supersymmetry breaking takes place non-perturbatively in the second vacuum, it 
is caused by the strong interactions. When $f_a(T,\, z)\simeq const$, the gauge 
couplings at high energies are almost identical in both vacua and their
running down to the SUSY breaking scale $M_{S}\sim m_{3/2}\sim m_{\alpha}$ are also 
the same. Then using the matching condition $\alpha^{(2)}_3(M_S)=\alpha^{(1)}_3(M_S)$, 
one finds the scale $\Lambda_{SQCD}$, where the supersymmetric QCD interactions 
become strong in the second vacuum:
\begin{equation}
\Lambda_{SQCD}=M_{S}\exp\left[{\frac{2\pi}{b_3\alpha_3^{(2)}(M_{S})}}\right]\,,\qquad
\frac{1}{\alpha^{(2)}_3(M_S)}=\frac{1}{\alpha^{(1)}_3(M_Z)}-
\frac{\tilde{b}_3}{4\pi}\ln\frac{M^2_{g}}{M_Z^2}-\frac{b'_3}{4\pi}\ln\frac{M^2_{S}}{M_g^2}\, .
\label{22}
\end{equation}
In Eq.(\ref{22}) $\alpha^{(1)}_3$ and $\alpha^{(2)}_3$ are the values of the strong gauge
couplings in the physical and second vacua respectively, $M_g$ is the gluino mass, while 
$\tilde{b}_3=-7$, $b_3=-3$ and $b'_3=-5$ are the one--loop beta functions for the strong gauge 
coupling in the SM, MSSM and Split SUSY scenario respectively. At the scale $\Lambda_{SQCD}$ 
the t--quark Yukawa coupling in the MSSM is of the same order of magnitude as the strong 
gauge coupling. The large Yukawa coupling of the top quark may result in the formation of 
a quark condensate that breaks supersymmetry, inducing a non--zero positive value for 
the cosmological constant $\Lambda \simeq \Lambda_{SQCD}^4$.

\begin{figure}
\centering
~\hspace*{-7.5cm}{$\log[\Lambda_{SQCD}/M_{Pl}]$}\\
\includegraphics[width=.6\textwidth]{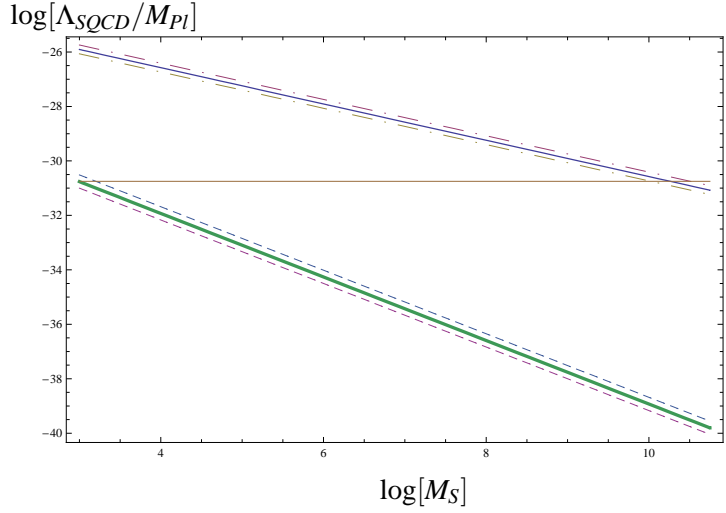}\\
~\hspace*{1cm}{$\log[M_S]$}\\
\caption{ The value of $\log\left[\Lambda_{SQCD}/M_{Pl}\right]$ versus $\log M_S$
for $M_q=M_g=500\,\mbox{GeV}$. The thin and thick solid lines correspond to the  
Split SUSY scenarios with the pure MSSM particle content and the MSSM particle   
content supplemented by an additional pair of $5+\bar{5}$ multiplets respectively.
The dashed and dash--dotted lines represent the uncertainty in $\alpha_3(M_Z)$.
The thin and thick solid lines are obtained for $\alpha_3(M_Z)=0.1184$,
the upper (lower) dashed and dash-dotted lines correspond to
$\alpha_3(M_Z)=0.116$ ($\alpha_3(M_Z)=0.121$). The horizontal line represents the
observed value of $\Lambda^{1/4}$. The SUSY breaking scale $M_S$ is measured in GeV.}
\label{lambda-split-susy}
\end{figure}

In Fig.~1 the dependence of $\Lambda_{SQCD}$ on the SUSY breaking scale $M_S$ is examined. 
We set $M_g=500\,\mbox{GeV}$. From Fig.~1 one can see that the value of $\Lambda_{SQCD}$ 
is much lower than the QCD scale in the Standard Model and diminishes with increasing $M_S$. 
The measured value of the cosmological constant is reproduced when $M_S\sim 10^{10}\,\mbox{GeV}$. 
With increasing gluino mass the value of the SUSY breaking scale that results in an appropriate 
value of the cosmological constant decreases. The results of our numerical analysis indicate 
that for $\alpha_3(M_Z)=0.116-0.121$ and $M_g=500-2500\,\mbox{GeV}$ the value of $M_S$ varies 
from $2\cdot 10^9\,\mbox{GeV}$ up to $3\cdot 10^{10}\,\mbox{GeV}$. 

The obtained prediction for the supersymmetry breaking scale can be tested. A striking 
feature of the Split SUSY model is the extremely long lifetime of the gluino. In the 
considered case the gluino decays through a virtual squark to $q\bar{q}+\chi_1^0$.
The large mass of the squarks then implies a long lifetime for the gluino. This lifetime
is given by 
\begin{equation}
\tau \sim 8\biggl(\frac{M_S}{10^9\,\mbox{GeV}}\biggr)^4 
\biggl(\frac{1\,\mbox{TeV}}{M_g}\biggr)^5\,s.
\label{241}
\end{equation}
If, as is predicted, the SUSY breaking scale lies in the interval from 
$2\cdot 10^9\,\mbox{GeV}$ ($M_g=2500\,\mbox{GeV}$) to 
$3\cdot 10^{10}\,\mbox{GeV}$ ($M_g=500\,\mbox{GeV}$)
the corresponding gluino lifetime lies in the interval from 
$1\,\mbox{sec.}$ to $2\cdot 10^8\,\mbox{sec.}$ ($1000\,\mbox{years}$). 
Thus the measurement of the gluino lifetime will allow an estimate to be made of
the value of $M_S$ in the Split SUSY model. 

The observed cosmological constant can also be reproduced for a much lower 
SUSY breaking scale, if the MSSM particle content is supplemented by an additional pair 
of $5+\bar{5}$ supermultiplets. In the physical vacuum new bosonic states 
associated with the extra $5+\bar{5}$ multiplets gain masses around $M_S$ while their 
fermionic partners can have masses of the order of the gluino mass. So we set
the masses of the new vectorlike quarks $M_q$ equal to $M_g$. Then from Fig.~1 it is 
easy to see that the observed value of the cosmological constant can be reproduced 
even for $M_S\simeq 1\,\mbox{TeV}$.


\begin{thebibliography}{99}
\bibitem{1}
C.~Froggatt, R.~Nevzorov and H.~B.~Nielsen,
\emph{On the smallness of the cosmological constant in SUGRA models},
\emph{Nucl.\ Phys.\  B} {\bf 743} (2006) 133 [{\tt hep-ph/0511259}].
\bibitem{2}
C.~Froggatt, L.~Laperashvili, R.~Nevzorov and H.~B.~Nielsen,
\emph{No-scale supergravity and the multiple point principle},
[{\tt hep-ph/0411273}].
\bibitem{3}
D.~L.~Bennett and H.~B.~Nielsen,
\emph{Predictions for nonAbelian fine structure constants from
multicriticality},
\emph{Int.\ J.\ Mod.\ Phys.\  A} {\bf 9} (1994) 5155 [{\tt hep-ph/9311321}].
\bibitem{31}
D.~L.~Bennett, C.~D.~Froggatt and H.~B.~Nielsen,
\emph{Nonlocality as an explanation for fine tuning and field replication in
nature}, [{\tt hep-ph/9504294}].
\bibitem{4}
C.~Froggatt, L.~Laperashvili, R.~Nevzorov and H.~B.~Nielsen,
\emph{Cosmological constant in SUGRA models and the multiple point principle},
\emph{Phys.\ Atom.\ Nucl.}  {\bf 67} (2004) 582 [{\tt hep-ph/0310127}].
\end{thebibliography}
\end{document}